\documentclass[epj,nopacs]{svjour}
\usepackage{graphics}
\usepackage{epsfig}
\usepackage{latexsym}
\newcommand{\spar}{{\stackrel{\rightarrow}{\Rightarrow}}}
\newcommand{\sant}{{\stackrel{\rightarrow}{\Leftarrow}}}

\begin{document}
\hugehead

\title{The $Q^2$--dependence of the generalised Gerasimov--Drell--Hearn
  integral for the deuteron, proton and
  neutron}

\author{
The HERMES Collaboration \medskip \\
A.~Airapetian,$^{32}$
N.~Akopov,$^{32}$
Z.~Akopov,$^{32}$
M.~Amarian,$^{26,32}$
V.V.~Ammosov,$^{24}$
E.C.~Aschenauer,$^{6}$
R.~Avakian,$^{32}$
A.~Avetissian,$^{32}$
E.~Avetissian,$^{10,32}$
P.~Bailey,$^{15}$
% ...
}

\institute{ 
$^1$Department of Physics, University of Alberta, Edmonton, Alberta T6G 2J1, Canada\\
$^2$Physics Division, Argonne National Laboratory, Argonne, Illinois 60439-4843, USA\\
$^3$Istituto Nazionale di Fisica Nucleare, Sezione di Bari, 70124 Bari, Italy\\
$^4$Nuclear Physics Laboratory, University of Colorado, Boulder, Colorado 80309-0446, USA\\
$^5$DESY, Deutsches Elektronen-Synchrotron, 22603 Hamburg, Germany\\
$^6$DESY Zeuthen, 15738 Zeuthen, Germany\\
$^7$Joint Institute for Nuclear Research, 141980 Dubna, Russia\\
$^8$Physikalisches Institut, Universit\"at Erlangen-N\"urnberg, 91058 Erlangen, Germany\\
$^9$Istituto Nazionale di Fisica Nucleare, Sezione di Ferrara and Dipartimento di Fisica, Universit\`a di Ferrara, 44100 Ferrara, Italy\\
$^{10}$Istituto Nazionale di Fisica Nucleare, Laboratori Nazionali di Frascati, 00044 Frascati, Italy\\
$^{11}$Fakult\"at f\"ur Physik, Universit\"at Freiburg, 79104 Freiburg, Germany\\
$^{12}$Department of Subatomic and Radiation Physics, University of Gent, 9000 Gent, Belgium\\
$^{13}$Physikalisches Institut, Universit\"at Gie{\ss}en, 35392 Gie{\ss}en, Germany\\
$^{14}$Department of Physics and Astronomy, University of Glasgow, Glasgow G128 QQ, United Kingdom\\
$^{15}$Department of Physics, University of Illinois, Urbana, Illinois 61801, USA\\
$^{16}$Physics Department, University of Liverpool, Liverpool L69 7ZE, United Kingdom\\
$^{17}$Department of Physics, University of Wisconsin-Madison, Madison, Wisconsin 53706, USA\\
$^{18}$Laboratory for Nuclear Science, Massachusetts Institute of Technology, Cambridge, Massachusetts 02139, USA\\
$^{19}$Randall Laboratory of Physics, University of Michigan, Ann Arbor, Michigan 48109-1120, USA \\
$^{20}$Lebedev Physical Institute, 117924 Moscow, Russia\\
$^{21}$Sektion Physik, Universit\"at M\"unchen, 85748 Garching, Germany\\
$^{22}$Nationaal Instituut voor Kernfysica en Hoge-Energiefysica (NIKHEF), 1009 DB Amsterdam, The Netherlands\\
$^{23}$Petersburg Nuclear Physics Institute, St. Petersburg, Gatchina, 188350 Russia\\
$^{24}$Institute for High Energy Physics, Protvino, Moscow oblast, 142284 Russia\\
$^{25}$Institut f\"ur Theoretische Physik, Universit\"at Regensburg, 93040 Regensburg, Germany\\
$^{26}$Istituto Nazionale di Fisica Nucleare, Sezione Roma 1, Gruppo Sanit\`a and Physics Laboratory, Istituto Superiore di Sanit\`a, 00161 Roma, Italy\\
$^{27}$Department of Physics, Simon Fraser University, Burnaby, British Columbia V5A 1S6, Canada\\
$^{28}$TRIUMF, Vancouver, British Columbia V6T 2A3, Canada\\
$^{29}$Department of Physics, Tokyo Institute of Technology, Tokyo 152, Japan\\
$^{30}$Department of Physics and Astronomy, Vrije Universiteit, 1081 HV Amsterdam, The Netherlands\\
$^{31}$Andrzej Soltan Institute for Nuclear Studies, 00-689 Warsaw, Poland\\
$^{32}$Yerevan Physics Institute, 375036 Yerevan, Armenia\\
}

\date{Received: \today / Revised version:}

\titlerunning{The $Q^2$--dependence of the gen. GDH integral for $p$ and $n$}
\authorrunning{The HERMES Collaboration}

\abstract{ The Gerasimov--Drell--Hearn (GDH) sum rule connects the anomalous
  contribution to the magnetic
  moment of the target nucleus with an energy--weighted integral of the difference
  of the helicity--dependent photoabsorption cross sections. Originally conceived for real
  photons, the GDH integral  can be generalised to the case of photons with
  virtuality $Q^2$. For spin--1/2 targets such as the nucleon, it then
  represents the non-perturbative limit of  the first moment $\Gamma_1$ of the spin structure function $g_1(x,Q^2)$
  in deep inelastic scattering (DIS).
The data collected by HERMES with a deuterium target are presented together
  with a re-analysis of previous measurements on the proton.
 This
  provides an unprecedented and complete
  measurement of the generalised GDH integral for photon--virtuality ranging over $1.2<Q^2<12.0$ ~GeV$^2$ and
  for photon--nucleon invariant mass squared $W^2$ ranging over 
  $1<W^2<45$ ~GeV$^2$, thus covering simultaneously the nucleon-resonance and the deep inelastic
  scattering regions. 
  These data allow the study of the $Q^2$--dependence of the full GDH integral, which is
  sensitive to both the $Q^2$--evolution of the resonance form factors and
  contributions of higher twist. The contribution of the nucleon-resonance region  is seen to decrease
  rapidly with increasing $Q^2$. The DIS contribution is 
  sizeable over the full measured range, even down to the lowest measured $Q^2$. As expected, at higher $Q^2$ the data are
  found to be  in
  agreement with previous measurements of the first moment of $g_1$. From data on the deuteron 
  and proton, the GDH integral for the
  neutron has been derived and the proton--neutron difference evaluated. This
  difference is found to satisfy  the fundamental Bjorken sum rule at $Q^2 = 5$~GeV$^2$.
} 
\maketitle


\section{Introduction}
The Gerasimov--Drell--Hearn (GDH) sum rule connects an energy--weighted integral of
the difference of the helicity--dependent real--photon absorption cross sections with
the anomalous contribution $\kappa = \frac{ \mu M_t}{eI}-Z$ to the magnetic moment $\mu$  of the target nucleus with atomic number $Z$~\cite{ger} (or nucleon~\cite{ger,dh}) :
\begin{equation}
\int_{\nu_0}^{\infty} \lbrack \sigma^{\sant}( \nu )-\sigma^{\spar}(
\nu ) \rbrack
\frac{d\nu }{\nu}=-\frac{4\pi^2 I \alpha }{M_t^2} \kappa^2.
\label{gdh}
\end{equation}
Here $\sigma^{\sant}$ and $\sigma^{\spar} $ are the photoabsorption cross sections for relative orientation   of the  photon spin anti--parallel and parallel  to the nucleus spin $I$, $\nu$ is the photon energy in the target rest frame, $\nu_0$
is the photoabsorption threshold, $M_t$ is the 
nucleus mass,  $\alpha$ the electromagnetic fine--structure constant and $e$ the elementary charge. This sum rule provides an
interesting link between the
helicity--dependent dynamics and a static ground state property of the target nucleus.

The GDH sum rule holds for any type of target, i.e. it is  valid for
protons, neutrons or nuclei. It is also considered to be important in
electroweak physics~\cite{electroweak}.
The GDH sum rule is
derived starting from the Compton forward--scattering
amplitude  following the general physics principles of Lorentz and gauge
invariance and is non--pertur\-ba\-tive in nature. The only questionable
assumption in its derivation is the use of an unsubtracted dispersion relation.
For the proton ($\kappa_{p}=+1.79$) the GDH sum rule prediction is
$-204\,\mu$b, for the neutron ($\kappa_{n}= -1.91$) it is 
$-233\,\mu$b. The prediction for the deuteron ($\kappa_{d}=-0.143)$ 
is $-0.65\,\mu$b. It should be noted that for nuclear targets the lowest--lying
inelastic channel is the break--up reaction, in contrast to photoabsorption on
the nucleon where the lowest--lying inelastic channel corresponds to single
pion production.

No test of the GDH sum rule was hitherto performed due to the lack of
polarised targets and suitable real--photon beams. Only recently, first results of
an experiment on polarised protons  in a limited beam energy range have 
been published~\cite{MAMIres}. Using extrapolations
into the unmeasured regions, Eq.~(\ref{gdh}) for the proton seems to be  satisfied within the
experimental uncertainties. Further real--photon experiments are underway at various laboratories 
to extend the energy range of the measurements~\cite{ELSA,otherexp}.

The GDH integral  can be generalised to non--zero photon virtuality $Q^2$
in terms of the helicity--dependent virtual--photon absorption cross sections $\sigma^{\sant}$ and $\sigma^{\spar} $~\cite{RP,drechsel}:

\begin{equation}\label{for1}
I_{GDH}(Q^2)=
\int_{\nu_{0}}^{\infty} \lbrack \sigma^{\sant}( \nu, Q^2 )-\sigma^{\spar}( \nu, Q^2 )
\rbrack
\frac{               
d\nu }{\nu}.
\end{equation}

The cross section difference appearing in the integrand is given by
\begin{equation}
\Delta \sigma = \sigma^{\sant} - \sigma^{\spar}= \frac{8\pi^2\alpha}{M_tK}\tilde{A_1}F_1.
\end{equation}
In terms of photon--nucleon (nucleus) helicity states this relation is valid for
any target; in case of the deuteron it comprises a mixture of vec\-tor and
ten\-sor states.  Here $\tilde{A_1}$ is the photon--nucleon (nucleus) helicity asymmetry, $F_1$ the unpolarised nucleon (nucleus) structure function and $K$ the virtual--photon flux factor.

Various generalisations of the GDH integral   have been considered in the
literature. The difference lies in the  choice made for
$K$. 
In the notation of Ref.~\cite{drechsel} three such generalisations were considered. In terms of  $\tilde{A_1}$ and $F_1$ they read:
\begin{eqnarray}
I_A(Q^2)&=&\frac{8 \pi^2 \alpha}{Q^2}\int_0^{x_0}\tilde{A_1}F_1dx, \label{ia}\\
I_B(Q^2)&=&\frac{8 \pi^2 \alpha}{Q^2}\int_0^{x_0}\frac{1}{\sqrt{1+\gamma^2}}\tilde{A_1}F_1dx,
\label{ib}\\
I_C(Q^2)&=&\frac{8 \pi^2 \alpha}{Q^2}\int_0^{x_0}\frac{1}{1-x}\tilde{A_1}F_1dx, \label{ic}
\end{eqnarray}
with $x=Q^2/2M\nu$.
$I_A$ corresponds to the case $ K=\nu$. The Gilman notation
$K = \nu \sqrt{1+\gamma^2}$ ~\cite{gil} has been used for $I_B$
while for $I_C$  the Hand convention $K = \nu(1-x)$ ~\cite{Hand} was chosen. 
They  all are numerically close to each other in the limits of deep inelastic scattering and
real--photon absorption, but lead to
different numerical results for intermediate $Q^2$. 
As was pointed out in Ref.~\cite{drechsel}, the generalisation given in
Eq.~(\ref{ib}) is most clearly related to photoabsorption cross sections. 
Hence, the generalisation used for the figures in this paper is $I_B$.
The full numerical results will be given for all three prescriptions.

When considering a nucleon target (spin $\frac{1}{2}$, mass $M$) the photon helicity asymmetry $\tilde{A_1}$ is
identical to the longitudinal virtual--photon asymmetry $A_1$ and the generalised GDH integral can be written in terms of the spin structure functions $g_1$ and $g_2$ as:
\begin{equation}
\label{for2}
I_{GDH}(Q^2) = \frac{ 8 \pi^2 \alpha}{M} \int_0^{x_{0}}\frac{g_1(x,Q^2)-\gamma^2 g_2(x,Q^2)}
{K}
\frac{dx}{x},
\end{equation}
where $g_1$ and $g_2$ are the polarised structure functions of the nucleon,
$\gamma^2=Q^2/\nu^2$,
$x_{0}=Q^2/2M\nu_{0}$.

Examining the generalised GDH integral provides a way to study the transition
from polarised real--photon absorption ($Q^2=0$) on the nucleon to polarised
deep inelastic lepton scattering (DIS). 
 In other words, it
constitutes an observable that allows the study of
the transition from the non--perturbative regime at low $Q^2$  to the perturbative regime at high $Q^2$. 
Since the generalised GDH integral  is calculated for inelastic reactions, elastic scattering is excluded from its calculation.
As has been pointed out in Ref.~\cite{JiMel}, the elastic
contribution to the photon cross section becomes the dominant one below $Q^2 \simeq 0.5$ $ \mbox{GeV}^2$; it has to be taken into
account when comparing with twist expansions of the first moment of the spin
structure function $g_1$.
In the kinematic region considered in this paper, elastic contributions are expected to be small.

Assuming that
the Burkhardt -- Cotting\-ham sum rule
\begin{equation}
\int_0^1 g_2(x,Q^2)dx=0
\end{equation}
holds in good approximation due to the relatively large $Q^2$ values considered in this paper, then Eq.~(\ref{for2}) simplifies to 
\begin{equation}\label{eq:dis}
I_{A}(Q^2) = \frac{16\pi^2\alpha}{Q^2}\Gamma_1(Q^2).
\end{equation}
As $Q^2$ becomes larger, the other generalisations $I_B$ and eventually $I_C$ also converge to this value.
The first moment of the spin structure function $g_1$,
$\Gamma_1=\int_0^1g_1(x)dx$, is predicted to have at large $Q^2$ only a logarithmic
$Q^2$ dependence from QCD evolution. 
Since for the proton $\Gamma_1^p >0$ for higher $Q^2$, $I_{GDH}^p$ must
change sign as $Q^2$ approaches zero in order to reach the negative value predicted by the GDH sum rule
at the real--photon point. The different generalisations lead to different values for the expected zero
crossing needed to connect the negative value predicted by Eq.~(\ref{gdh}) with
the positive value required by measurements of $\Gamma_1^p$ in the DIS
limit.  
 For the neutron  $\Gamma_1^n$ is negative for all measured $Q^2$.

The difference of the GDH integral for the proton and the neutron, $I_{GDH}^p -
I_{GDH}^n$, is of
great interest. In the real--photon case, the GDH sum rule gives $I_{GDH}^p -
I_{GDH}^n = 29 \mu$b, with  a
sign  opposite to what results from multipole analyses of meson photoproduction
data~\cite{RP567}. 
In the Bjorken limit  the difference $\Gamma_1^p-\Gamma_1^n$ is given by the
Bjorken sum rule. It can be
  derived using only current algebra and isospin symmetry~\cite{bjorken}.
This sum rule relates the difference of the first moments of
 $g_1^p$ and $g_1^n$ at fixed $Q^2$ to the well--measured neutron beta--decay
coupling constant $g_a=|g_A/g_V|=1.2670\pm0.0035$ \cite{PDG}:
\begin{equation}\label{eq:bjorken}
\Gamma_1^p-\Gamma_1^n=\frac{1}{6}\cdot g_a\cdot C_{ns}(\alpha_s(Q^2)),
\end{equation}
where $C_{ns}$ is the non--singlet QCD correction calculated thus far up to
${\cal{O}}(\alpha_s^3)$ in the modified
minimal subtraction ($\overline{MS}$) scheme \cite{lar2}. 
Experimental verification  of the Bjorken sum rule at finite $Q^2$ provides a fundamental test
of QCD. A measurement of $I_{GDH}^p -
I_{GDH}^n$ at large enough $Q^2$ provides such a test. Previous measurements
are consistent with
 the sum rule  when perturbative QCD corrections are
included \cite{SLAC,E155,SMC}.

The $Q^2$--dependence of the generalised GDH integral can be studied
separately in the DIS region, characterised by large photon--nucleon
invariant mass squared $W^2=M^2+ 2M\nu -Q^2$, and in the nucleon--resonance region
where $W^2$ amounts  to only a few GeV$^2$. 
Several experiments  measure the generalised GDH integral 
at low and intermediate
  $Q^2$, but cover kinematically only the low--$W^2$ region~\cite{GDHJlab,CLAS,hallA}. On the other hand, the high--$W^2$
  contribution to the generalised GDH integral is found to be sizeable and
  essential to any estimate of the total integral~\cite{gdhdis,gdhres}. Preliminary data  from real--photon
  experiments at
  higher energies support this statement~\cite{ELSA}. The kinematics of the
  HERMES experiment allow the study of the $Q^2$--development of the generalised
  GDH integral simultaneously in both the nucleon-resonance and DIS regions.

In section~\ref{sec:exp} the experimental setup for data taken with a deuteron
(proton) target will be described followed by a description of the analysis
procedure for both targets in section~\ref{sec:ana}. The results for the
deuteron nucleus are presented in section~\ref{sec:deutres} together with the
proton data re--analysed with respect to Ref.~\cite{gdhres} using an updated value for the target polarisation. From these two data sets the value of $I_{GDH}^n$ is calculated in section~\ref{sec:neutres}. Here the assumption is made that in the kinematical range under consideration, nuclear effects are small and the deuteron can be treated as consisting of two quasi--free nucleons. The results on the deuteron nucleus, the proton and the neutron are discussed in section~\ref{sec:disc}. From the values on the proton and neutron, the proton--neutron difference is calculated and compared to the Bjorken sum rule prediction in section~\ref{sec:bsr}. A summary of the paper is given in section~\ref{sec:sum}.

\section{Experiment}\label{sec:exp}

HERMES data on the deuteron target were taken in 1998 to 2000 with a 27.57 ~GeV beam of
longitudinally polarised posi\-trons incident on a longitudinally polarised
atomic Deuterium gas target internal to the HERA storage ring at DESY. Data on
the proton were taken in 1997  using a longitudinally polarised
atomic Hydrogen target.
The lepton beam polarisation was measured continous\-ly using Compton
backscattering of circularly polarised laser light~\cite{bar,beck}. The
average  beam polarisation for the deuteron (proton) data set was 0.55 (0.55)
with a fractional systematic uncertainty of 2.0\% (3.4\%).
 
The HERMES polarised gas target~\cite{ste} consists of polarised atomic D (H) 
confined in a storage cell.
It is fed with nuclear--polarised atoms by an atomic--beam source based on
Stern--Gerlach separation~\cite{sto} and  provides an areal target density of about $
2\times 10^{14} \; (7\times 10^{13})$ atoms/cm$^2$. The nuclear polarisation of atoms and the atomic
fraction are continously measured with a Breit--Rabi polarimeter~\cite{BRP} and
a target gas analyser~\cite{TGA}, respectively. The polarisation of the atoms can be flipped
within short time intervals providing both vector--polarisation states
and thus minimising systematic effects in spin--asymmetry mea\-sure\-ments.
The average value of the target polarisation for the deuteron (proton)
data was 0.85 (0.85) with a fractional systematic uncertainty of 3.5
(3.8)\%. The value of the proton target polarisation used for the data
presented in this paper has been updated with respect to Ref.~\cite{gdhres} making use of improved knowledge of sampling
corrections and treatment of molecular polarisation~\cite{targetpaper}. The luminosity was
monitored by detecting Bhabha events using calorimeter detectors close to the beam pipe~\cite{lumi}. The integrated
luminosity per nucleon of the deuteron (proton) data set was 222 pb$^{-1}$ (70 pb$^{-1}$).

Scattered positrons, as well as  coincident hadrons, were detected by the
HERMES spectrometer~\cite{hspect}. Positrons were 
distinguished from hadrons with an
average efficiency of 99\% and a hadron contamination of less than 1\% using
the information from an electromagnetic calorimeter, a transition--radiation
detector, a preshower scintillation counter and a Cherenkov counter. Only
the information on the scattered positron was used in this analysis.

\section{Data analysis}\label{sec:ana}

In the following, the analysis procedure used for the deu\-te\-ron
data is given. The analysis
procedure and treatment of  systematic uncertainties have been taken from Refs.~\cite{gdhdis,g1p}
and are detailed in Ref.~\cite{gdhres}, where the same analysis for the proton data
was performed. For completeness, the values and parametrisations used in the latter 
are given below. Note that, compared to Ref.~\cite{gdhdis}, the proton data 
set has been re--analysed in the full kinematic range of Ref.~\cite{gdhres}
to optimise the binning of the kinematically more restricted
nucleon-resonance region, where the detector acceptance prevents the full coverage
over $Q^2$.

The kinematic requirements imposed on the scattered positrons in the analysis
were identical for both targets. The full range in $W^2$ ($1.0<W^2<45$ ~GeV$^2$)
was separated into nucleon resonance region ($1.0<W^2<4.2$ ~GeV$^2$)
and  DIS region ($4.2<W^2<45.0$ ~GeV$^2$).
The $Q^2$-range $1.2<Q^2<12.0$ ~GeV$^2$ was divided into six bins;
the same  binning as in the proton case was chosen for the analysis of the
deuteron data and for the subsequent determination of $I_{GDH}^n$.
After applying data quality criteria, 0.55 (0.13) million events on the deuteron (proton) in the
nucleon-resonance region and 8.3 (1.4) million events in the DIS region were selected. 

For all  positrons detected, the angular resolution was better than 0.6 mrad,
the momentum resolution (aside from Bremsstrahlung tails) better than 1.6\% 
and the $Q^2$--resolution better
than 2.2\%. The threshold Cherenkov detector used in the proton measurement was replaced
by a Ring--imaging Cherenkov detector~\cite{RICH} for the data taking on the deuteron.
The additional amount of material led to a slightly worse $W^2$--resolution of
$\delta W^2 \approx 1.0 $ GeV$^2$ for the deuteron as compared to the proton
measurement ($\delta W^2 \approx 0.82 $ GeV$^2$) .
 Although these  $W^2$--resolutions do not allow distinguishing individual nucleon resonances, 
the integral measurement in the nucleon-resonance region is not degraded. 
 
The generalised GDH integral Eq.~(\ref{for1}) can be re--written  for any
target in terms of the
photon--target helicity asymmetry $\tilde{A_1}$ and the unpolarised structure function $F_1$: 
\begin{equation}\label{gdhexp}
I_{GDH}(Q^2)=\frac{8 \pi^2 \alpha}{M_t}\int_0^{x_{0}}\frac{\tilde{A_1}(x,Q^2) F_1(x,Q^2)}
{K}
\frac{dx}{x},
\end{equation}
where $K$ is the virtual--photon flux factor.

 The cross--section asymmetry $\tilde{A_1}$ for the absorption
of virtual photons was calculated from the measured cross section asymmetry $A_\parallel$  as
\begin{equation}\label{a1}
\tilde{A_1} = \frac{A_\parallel}{D} - \eta \tilde{A_2}.
\end{equation}
For spin--$\frac{1}{2}$ targets the photon helicity asymmetry $\tilde{A_1}$ is identical to the longitudinal virtual photon asymmetry $A_1$ and $\tilde{A_2}$ is identical to $A_2$. The difference between these two asymmetries is relevant for the deuteron target only. Even here it is considered to be small in the kinematic region examined in this paper and hence will be neglected in the following.

 The measured cross section asymmetry $A_\parallel$ is given by
\begin{equation}
A_\parallel = \frac{N^{\sant}L^{\spar}-N^{\spar}L^{\sant}}{N^{\sant}L^{\spar}_P+N^{\spar}L^{\sant}_P}.
\end{equation}
Here $N$ is the number of detected scattered positrons, $L$ is the integrated
luminosity corrected for dead time and $L_P$ is the integrated luminosity
corrected for dead time and weighted by the product of the beam and target
polarisations. The superscript $\spar(\sant)$ refers to the orientation of the target
spin parallel (anti--parallel) to the positron beam polarisation.
The kinematic factor $\eta$ is given by
\begin{equation}
\eta  = \frac{\gamma(1-y-\gamma^2y^2/4)}{(1-y/2)(1+\gamma^2y/2)},
\end{equation}
 where $y = \nu/E_{beam}$ is the inelasticity of the reaction.
The effective polarisation of the photon $D$ 
\begin{equation}
D  =  \frac{y(2-y)(1+\gamma^2y/2)}{y^2(1+\gamma^2)(1-2m_e^2/Q^2)+2(1-y-\gamma^2y^2/4)(1+R)} 
\end{equation}
depends also  on $R=\sigma_L/\sigma_T$, the ratio of the
absorption cross sections for longitudinal and transverse virtual
photons and  the electron mass $m_e$. $A_2$ is related to
lon\-gi\-tu\-di\-nal--\-trans\-verse photon--nucleon 
interference  and is not measured in the present experiment. 
In the DIS region $A_2$ can be parametrised in a general  form
as $A_2=cMx/\sqrt{Q^2}$, where $c$ is a constant determined from
a fit to the data given in Refs.~\cite{SLAC,E155} 
as $c=0.20 \; (0.53)$ for the deuteron (proton).
In the nucleon-resonance region  no data are available for the deuteron and $A_2=0$ was
chosen, while 
for the proton a constant value of $A_2=0.06\pm 0.16$   was adopted  as
obtained from SLAC measurements at $Q^2 = 3 \mbox{ GeV}^2$~\cite{SLAC}.

Radiative effects for both targets were calculated using the codes described in
Ref.~\cite{polrad}.  They were found not to exceed 7\% (4\%) of the
asymmetry $A_1$ for the deuteron (proton). On the integral level they do not
exceed 2\% and were included in the systematic uncertainty.

\begin{figure}
\centerline{\epsfig{figure=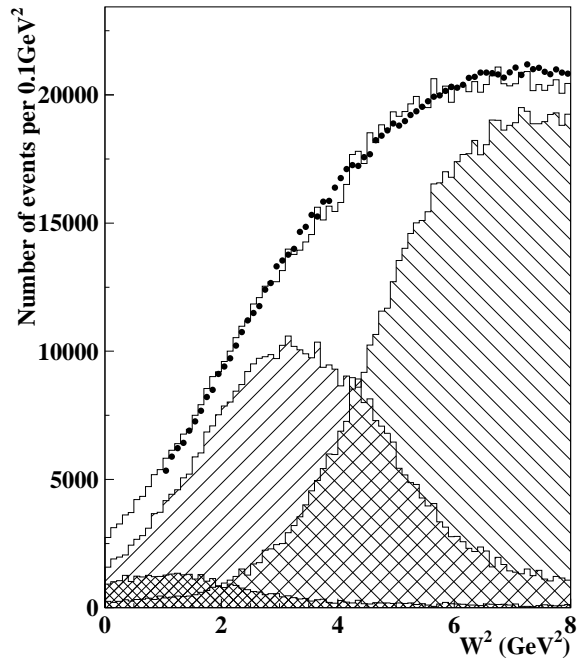,width=8.0cm}}
\caption{Comparison of deuteron data with Monte Carlo simulations  for the nucleon-resonance region as a function of $W^2$. The total simulated distribution has been normalised to the data. 
  The cross--hatched area represents the contribution from quasi--elastic 
  scattering, while the lined areas show the contribution from the nucleon-resonance
  region (left) and from the DIS region (right). The
  solid line indicates the sum of all simulated events and compares favourably with 
  the data points. The statistical uncertainties of the data are covered by the symbols.}
\label{fig:mocaplot}       
\end{figure}

 The fraction of events smeared from the DIS to the nucleon-resonance region and vice versa is 
evaluated by a Monte Carlo simulation of both regions including  radiative and
detector effects. 
Smearing effects in the deep inelastic region have been
evaluated for all targets following the procedures described in
Ref.~\cite{gdhres}.
The events on the deuteron (proton) were simulated using the parametrisation of $F_2$ from 
Ref.~\cite{nmcd8} (\cite{nmc}) for the DIS  region, the elastic form factors from
Ref.~\cite{stein}(\cite{bilen}) and the parametrisation of $F_2$ in  the nucleon-resonance region from
Ref.~\cite{bodekneut} for both targets. Fig.~\ref{fig:mocaplot} shows the distribution of experimental
data as a function of $W^2$ in comparison with the simulated events on the
deuteron. It is apparent that the shape of the simulated distribution agrees
well with the data. Similar agreement  has been found for the proton.

For the deuteron (proton) case, the relative contaminations from the quasi--elastic (elastic) and deep
inelastic region in the nucleon-resonance region range from 15\% (10\%) to 3\% (2\%)
and from 11\% (7\%)
to 23\% (16\%) respectively, as $Q^2$ increases from 1.2 ~GeV$^2$ to 12.0 ~GeV$^2$. 
The fraction of events smeared from the nucleon-resonance region to the deep inelastic region
ranged from 2.9\% (2.5\%) to 0.5\% (0.2\%), respectively. Smearing from the 
elastic region to the DIS region can be neglected in the present experiment.

To evaluate the systematic uncertainty from smearing, two different assumptions
on $A_1$ for the deuteron (proton) have been used: a polynomial representation $A_1=-0.0307+0.92x-0.28x^2$ (power law $A_1=x^{0.727}$) that smoothly extends
the DIS behaviour for the asymmetry into the nucleon-resonance region ~\cite{nagaitsev};
and for both targets a step function ($A_1=-0.5$ for $W^2 < 1.8$~GeV$^2$ and $A_1=+1.0$ for
$1.8 $~GeV$^2 < W^2 < 4.2$~GeV$^2$ ) that is suggested by the hypothesis of the possible
dominance of the $P_{33}$--resonance at low $W^2$ and of the $S_{11}$--resonance
at higher $W^2$ (see e.g. Ref.~\cite{ede}). 
The combined systematic uncertainty in the partial integrals from smearing and radiative effects does
not exceed 14\% (10\%) for the deuteron (proton) data. In both cases,
smearing gives by far the dominant contribution. 

\section{Results for Deuteron and Proton}\label{sec:deutres}

The GDH integrals for the deuteron and proton were evaluated following the
procedure described in the previous section. The nucleon--resonance region and the DIS
region were treated sep\-a\-rate\-ly. The large $W^2$--range covered by the HERMES experiment
allows essentially the first experimental determination
of the complete generalised GDH integral for the deuteron, proton and neutron.
\begin{figure}
\centerline{\epsfig{figure=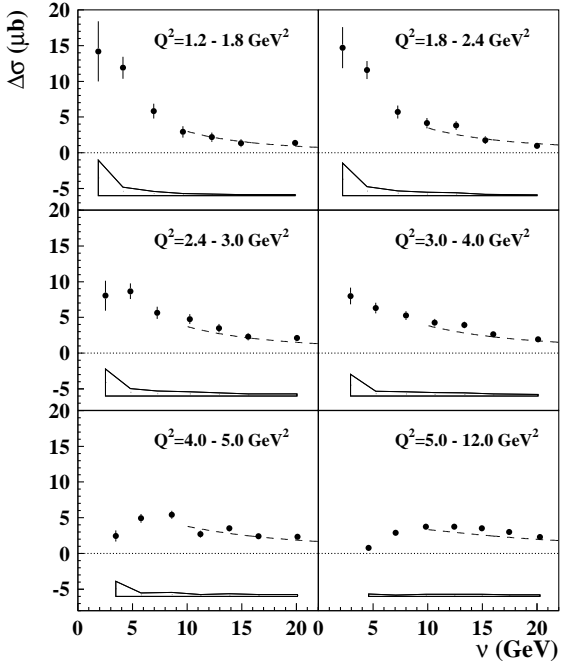,width=8.0cm}}
\caption{The cross section difference $\Delta \sigma$ obtained for the
  deuteron nucleus as a function of the virtual photon energy $\nu$ for
  various bins in $Q^2$. The dashed curves represent the parametrisation used to extrapolate into the unmeasured region at high $W^2$.}\label{fig:sdifd}
\end{figure}

The GDH integral $I_{GDH}^d$ for the deuteron  was evaluated 
using Eq.~(\ref{gdhexp})
 in both the nucleon-resonance region  and the DIS region. Here and in the following $I_{GDH}^d$ is understood as the generalised GDH integral for the deuteron {\em nucleus}. The unpolarised structure function
$F_1^d=F_2^d(1+\gamma^2)/(2x(1+R^d))$ was calculated in the nucleon--resonance region from a modification of the 
 parametrisation of $F_2^d$ given in Ref.~\cite{bodekneut} that accounts for nucleon
 resonance excitation assuming  $R^d=\sigma_L/\sigma_T$ to be constant and equal
 to 0.18 in the whole $W^2$--range.
In the DIS
region $F_1^d$ for the deuteron  was calculated
following a parametrisation of $F_2^d$ from Ref.~\cite{nmcd8}. In the same
kinematic region $R$ was chosen according to a fit in Ref.~\cite{whi}.
 Note that due to cancellations between
 the $R^d$ dependences of $F_1^d$ and $D$ at low $y$ the final result is
 affected by at most 2\% by a particular choice of $R^d$. The $W^2$--dependence of the
 integrand $F_1^d/K$  in the individual bins was fully accounted for in
 the integration.

\begin{figure}
\centerline{\epsfig{figure=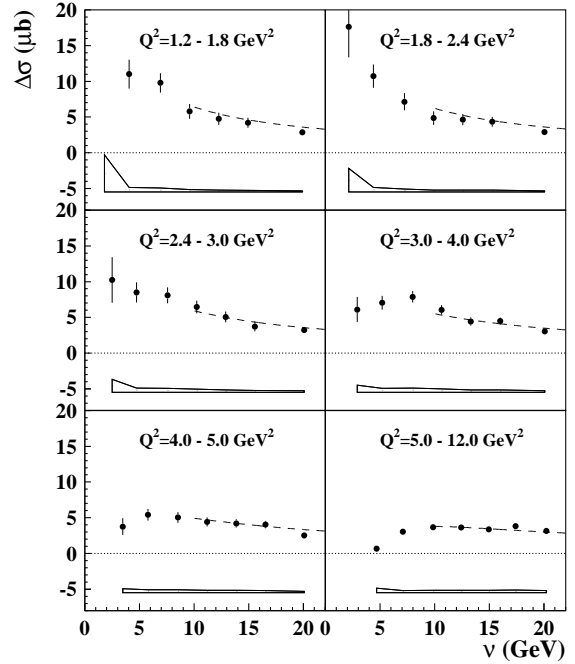,width=8.0cm}}
\caption{The cross section difference $\Delta \sigma$ obtained for the
  proton target as a function of the virtual photon energy $\nu$ for
  various bins in $Q^2$. Note that for the upper left plot the first data
  point corresponding to $28.18 \pm 6.79 \; \mu b $ at $\nu=1.8$~GeV is off scale. The dashed curves represent the parametrisation used to extrapolate into the unmeasured region at high $W^2$.}\vspace{-0.5cm}\label{fig:sdifp}       
\end{figure}

The integrand  $\Delta \sigma$ used to calculate  $I^d_{GDH}$ for the deute\-ron target is
  shown in Fig.~\ref{fig:sdifd} as a function of $\nu$ for the various bins
  in $Q^2$. For the proton case the corresponding values for
$\Delta \sigma$   are shown in Fig.~\ref{fig:sdifp}. In both cases, the extrapolation into the unmeasured region for $W^2>45$~GeV$^2$ was done
using a multiple--Reggeon exchange parametrisation~\cite{nikolo} for
$\Delta \sigma$ at high energy.  
The resulting contributions are given in Table~\ref{tab:results1} and range
  for the deuteron from -0.07 $\mu b$ at
$Q^2 = 1.5 $ GeV$^2$ to 1.53 $\mu b$ at $Q^2 = 6.5 $ GeV$^2$. The corresponding
contributions for the proton amount to about 3.5 $\mu
b$ for all $Q^2$--bins.

\begin{figure}
\centerline{\epsfig{figure=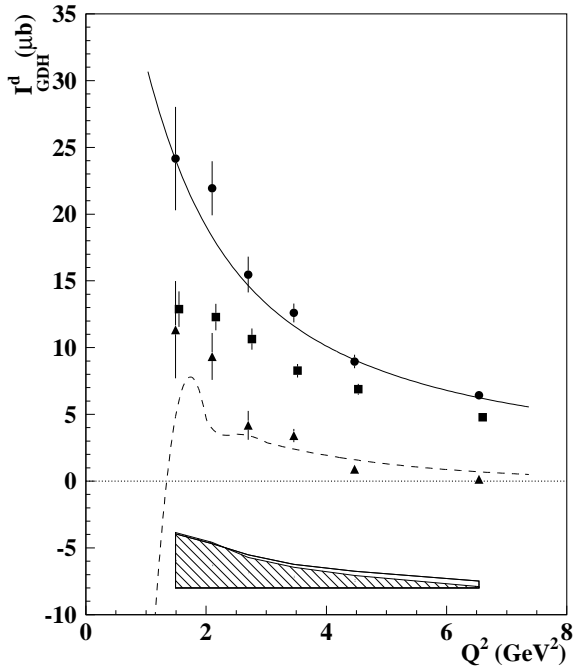,width=8.0cm}}
\caption{The generalised GDH integral $I_{GDH}^d$ for the deuteron nucleus, shown as a
  function 
  of $Q^2$ for the three  kinematic regions considered: nucleon-resonance region (triangles),
  DIS region (squares), and full $W^2$--region (circles) including extrapolation to the
  unmeasured part. The error bars show the
  statistical uncertainties.  The solid curve is taken from
  Ref.~\cite{teryaev} and represents a prediction for the full $I_{GDH}^d$. The dashed curve represents a model for the nucleon-resonance region from Ref.~\cite{azna}. The systematic uncertainties of the full integral are given as a
  band; the hatched area inside represents the systematic uncertainty of the
  nucleon-resonance region alone. Note that some data points are slightly shifted for
  better visibility.}
\label{fig:deutful}       
\end{figure}

The generalised GDH integrals for the deuteron data, calculated in the nucleon-resonance
region, in the DIS region and over the full $W^2$--range, are depicted 
in Fig.~\ref{fig:deutful}.
The statistical and systematic uncertainties of the full
$I_{GDH}$ are clearly dominated by the uncertainties in the nucleon-resonance
region. They are particularly large due to the smallness of $D$ and the large
size of $\eta$ accentuating the uncertainties in $A_2^d$, which
  amounts to 30\% of the nucleon-resonance contribution. The systematic uncertainty on $A_2^d$ in the DIS region does not contribute
significantly.
The systematic uncertainty for the extrapolation to the unmeasured region at
high $W^2$ of 5\% has been taken into account.
Further
sources of systematic uncertainties include the beam and target polarisations (5.5\%), the
spectrometer geometry (2.5\%), the combined smearing and radiative effects
(14\% of the partial integrals) and the
knowledge of $F_2$ (5\%). The total systematic uncertainty of the
  total GDH integral ranges  from 16\% at $Q^2=1.5$~GeV$^2$ to 7.5\% at
  $Q^2=6.5$~GeV$^2$. For the systematic uncertainties of the
  nucleon-resonance and DIS regions, independent sources of systematic uncertainties were
  added in quadrature, while the systematic uncertainties stemming from smearing
  effects and the knowledge of $F_2^d$ were added linearly. Only smearing
  effects from the quasi--elastic region to the measured range of $W^2$ had to be  taken
  into account for the total integral, thus reducing considerably its systematic uncertainty due to smearing  compared to the integrals calculated separately in the nucleon-resonance and DIS
  regions.

\begin{figure}
\centerline{\epsfig{figure=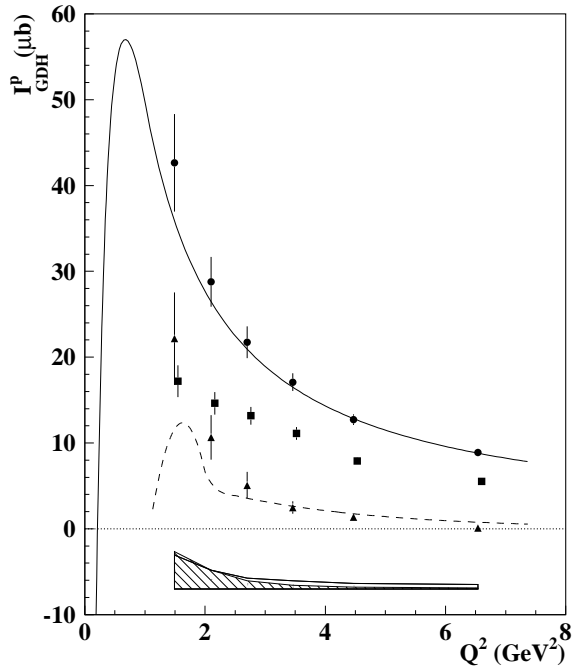,width=8.0cm}}
\caption{The generalised GDH integral $I_{GDH}^p$ for the proton using the same notation
  as in  Fig.~\ref{fig:deutful}. The solid curve is taken from
  Ref.~\cite{teryaev}, the dashed curve follows the model of Ref.~\cite{azna}. The data were   published earlier in
  Ref.~\cite{gdhres}, but are re--analysed for the present paper using improved
  knowledge on the target polarisation. The band representing the systematic
  uncertainties is given for the convention defined in Fig.~\ref{fig:deutful}. }\vspace{-0.5cm}\label{fig:proton}       
\end{figure}       

The generalised GDH integrals for the proton data reanalysed using an updated
value of the target polarisation, calculated in the nucleon-resonance
region, in the DIS region and over the full $W^2$--range, are depicted 
in Fig.~\ref{fig:proton}.
For both targets, the contribution of the nucleon-resonance region decreases
fas\-ter than that of the DIS
region as a function of $Q^2$. The latter dominates $I_{GDH}^d$ and $I_{GDH}^p$ for $Q^2>3.0$
~GeV$^2$ and remains sizeable even at the lowest measured $Q^2$.
The nucleon-resonance contribution shown in
Fig.~\ref{fig:deutful} and Fig.~\ref{fig:proton} respectively for the deuteron and proton is compared to a
curve representing the  prediction of the model of Ref.~\cite{azna}. This
model is based on a helicity--dependent sum over the first, second and third
nucleon--resonance regions using experimental resonance parameters, but assuming
infinitely narrow resonances. The threshold region was taken into account in
the first nucleon--resonance region. 
Within the total experimental uncertainties this model describes the data.

The data for the full integral on both the deuteron and the proton target
are compared to a model based on the leading twist
$Q^2$--evolution of the first moments of the two
polarised structure functions $g_1$ and $g_2$ without consideration of any
explicit nucleon--resonance contribution \cite{teryaev,tertwo}. In this model,  the
low--$Q^2$ behaviour of $g_2$  is governed by the $Q^2$--dependence of a
linear combination of the electric and magnetic Sachs form factors. The model predicts the shape. It
predictions thus depend on the experimental value for $\Gamma_1$ at
asymptotically large $Q^2$ and on the $Q^2$--dependence of the Sachs form
factors at low $Q^2$. The normalisations for asymptotically large $Q^2$ was taken from the present data. They were evaluated from the $Q^2$--dependencies of $I_{GDH}^{p,n}$ with the $1/Q^2$--dependence expected from leading twist devided out. 
Fitted by straight lines the results are $\Gamma_1^p=0.129 \pm 0.006$ and $\Gamma_1^n=-0.030\pm 0.007$ (cf. Fig.~{\ref{fig:neutq2}) . The parameterisation of the form factors was taken from Ref.~\cite{teryaev}. The model describes
the data on the proton very well. No explicit prediction for the deuteron is given; thus
the deuteron is modelled as the sum of proton and neutron. Nevertheless, the model prediction also  agrees well the deuteron data.

\section{Neutron results from Deuteron and Proton}\label{sec:neutres}

The extraction of the generalised GDH integral for the neutron from data taken on the deuteron and the proton requires nuclear effects such as Fermi motion and the depolarising effect of the D--state to be taken into account. These questions were addressed in  Ref.~\cite{ciofi}.
Following their model, the integral $I_{GDH}^n$ for the neutron was calculated from the results $I_{GDH}^d$
on the deute\-ron, as obtained in this analysis, and those on the
proton $I_{GDH}^p$ re--analysed following the procedure detailed in
Ref.~\cite{gdhres}:
\begin{equation}\label{eq:ciofi}
I_{GDH}^n=\frac{I_{GDH}^d}{1-1.5\omega_d}-I_{GDH}^p.
\end{equation}
 Here $\omega_d=0.050\pm 0.010$~\cite{lacombe} is the probability of the deuteron to be in a
 D--state.  It has
 been shown in Ref.~\cite{ciofi} that although the uncertainties in the structure
 functions in the integrand of Eq.~(\ref{gdhexp})  may be large, the resulting
 contribution  to the systematic uncertainty for
 the integral $I^n_{GDH}$ due to nuclear effects does
 not exceed 3\%.  This combined with the uncertainty in $\omega_d$
  leads to an additional systematic uncertainty on $I_{GDH}^n$ of 4\%.
No further assumptions, in particular not on $F_2^n$, $A_2^n$ and
$R$, are needed to derive the generalised GDH integral for the 
neutron using Eq.~(\ref{eq:ciofi}).

\begin{figure}
\centerline{\epsfig{figure=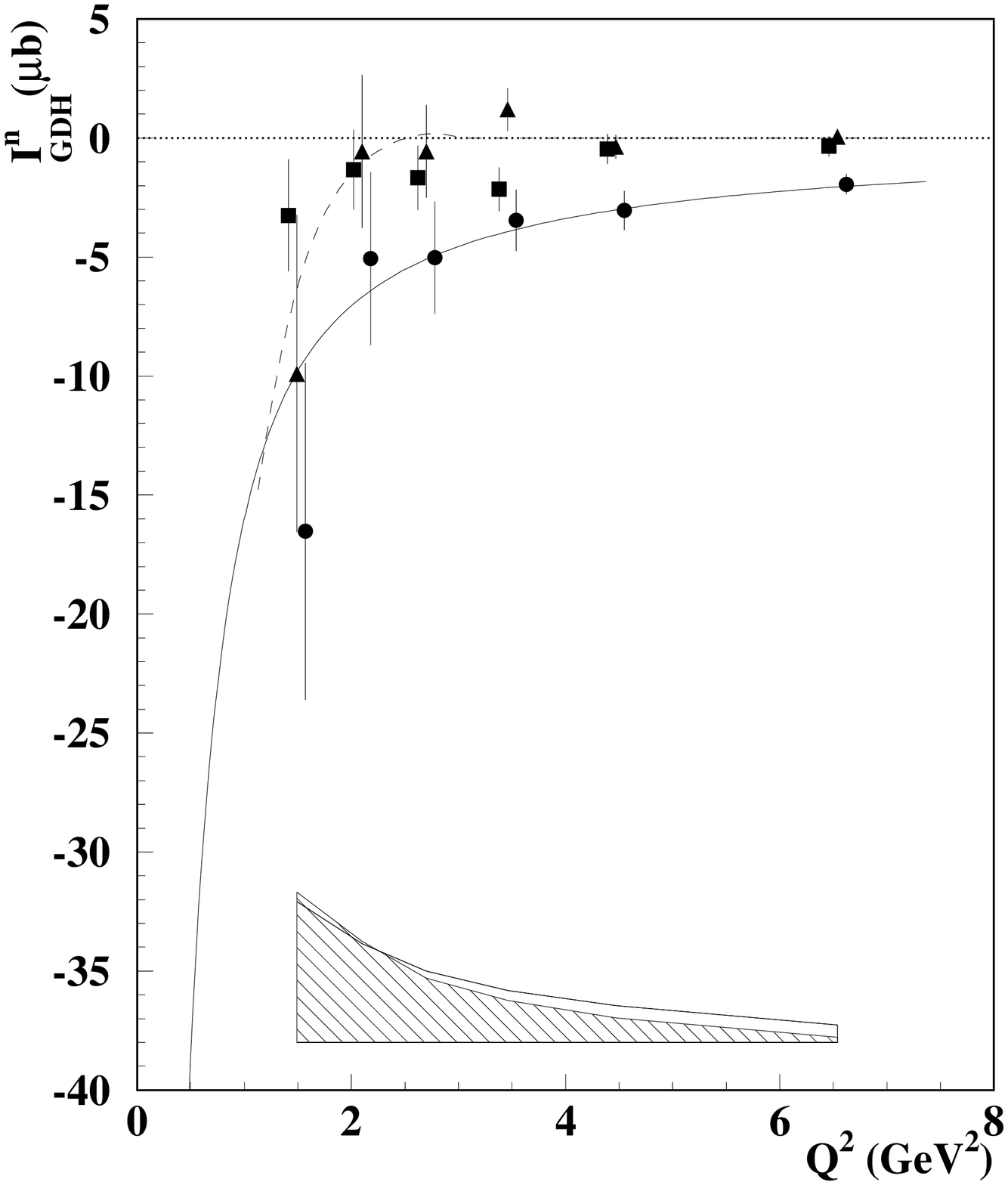,width=8.0cm}}
\caption{The generalised GDH integral $I_{GDH}^n$ for the neutron obtained
  from the deuteron and proton using the same notation
  as in Fig.~\ref{fig:deutful} for the symbols and theoretical curves. }
\label{fig:neutdis}       
\end{figure}

In the real--photon case, the application of Eq.~(\ref{eq:ciofi}) is not
straightforward, since  significant contributions from
 photodisintegration and coherent photoproduction must be taken into account~\cite{arenhoevel}. For virtual photons three different regions can be distinguished. For $Q^2 > 1 \mbox{ GeV}^2$ the generalised GDH integral can be described by the spin structure functions $g_1$ and $g_2$ taking twist--2 and twist--3 contributions into account (cf. Eq.~(\ref{for2})). Higher twist contributions are suppressed by powers of $1/Q^2$; Eq.~(\ref{eq:ciofi}) holds. For $ Q_0^2 < Q^2 < 1 \mbox{ GeV}^2$, Eq.~(\ref{eq:ciofi}) holds, but the GDH--integrals for proton and neutron deviate substantially from the $g_1$ and $g_2$ (twist--2 and twist--3) contributions. In this region, the Operator Product Expansion has already broken down and one has to resort to model assumptions like those of Refs.~\cite{teryaev,tertwo}. The relevant scale $Q_0^2$ was estimated in Ref.~\cite{terthree} to be $Q_0^2 \sim m_\pi^2$. Finally, for $Q^2 < Q_0^2$ Eq.~(\ref{eq:ciofi}) is no longer valid.

In Fig.~\ref{fig:neutdis} the results for $I_{GDH}^n$ obtained from the
 deu\-te\-ron and proton data in three $W^2$--regions are shown 
 together with  model predictions following Ref.~\cite{azna} and
 Ref.~\cite{teryaev}.
  As in  the proton case, the contribution from the nucleon resonance region
  decreases faster with increasing $Q^2$ compared to the contribution from the
 DIS region. The data are well described by the resonance model. The 
 contribution from the extrapolation to high $W^2$ is dominant for $Q^2 > 2.0$ ~GeV$^2$ and remains
 sizeable down to the lowest measured $Q^2$ (cf. Table~\ref{tab:results1}). In agreement with measurements of
 the neutron spin structure function $g_1^n$ and as expected from recent measurements of polarised
 quark distributions~\cite{deltaq}, $I_{GDH}^n$ is negative and of smaller absolute size
 than the proton value.
Within the total experimental uncertainties, the model prediction of
Ref. \cite{teryaev} agrees  well with the neutron data for the full integral.

Results on $I_{GDH}^n$ were also obtained from a previous measurement on a $^3$He target~\cite{gdhdis}. The neutron asymmetry was obtained from the $^3$He asymmetry
taking into account nuclear effects, the relative polarisation of the neutron and two
protons, as well as a fit to the data for $A_1^p$. Note that the lower $W^2$--limit for the data taken on $^3$He was 4 GeV$^2$ and thus slightly different
from the cut at 4.2 GeV$^2$ used in the deuteron analysis. Both data sets are
shown in Fig.~\ref{fig:neutcomp} and  agree within their respective uncertainties.
\begin{figure}
\centerline{\epsfig{figure=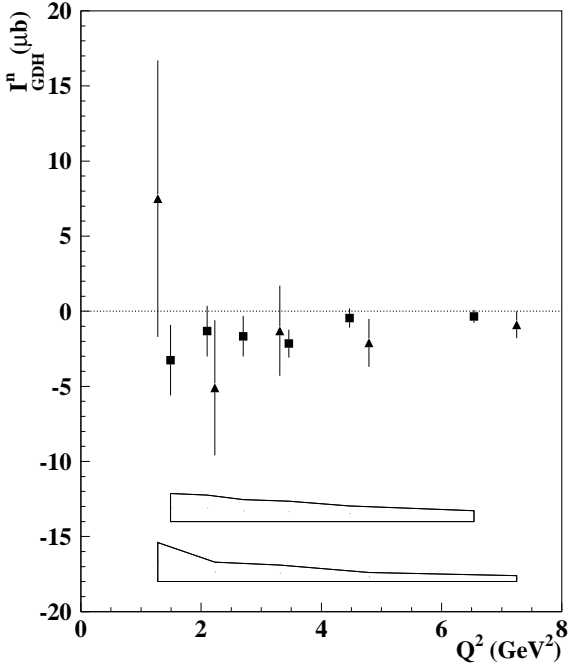,width=8.0cm}}
\caption{The generalised GDH integral $I_{GDH}^n$ in the DIS region for the neutron (squares)
  obtained combining deuteron and proton data, shown in 
  comparison to the results obtained on $^3$He (triangles)~\cite{gdhdis}. The lower
  $W^2$--limit for the latter was $W^2>4.0$~GeV$^2$ while for the former
  $W^2> 4.2$~GeV$^2$ was used.
   The error bars 
  represent the statistical uncertainties. The
  systematic uncertainties for each data set are given as error bands
  (deuteron top, $^3$He  bottom).}
\label{fig:neutcomp}       
\end{figure}

\section{Discussion of results} \label{sec:disc}
In Table~\ref{tab:results} the final results  are presented for the
full generalised GDH integrals on the deuteron, the proton and the neutron
in bins of $Q^2$ and for the three generalisations
(Eqs.~\ref{ia}, \ref{ib}, \ref{ic}) considered in the literature.
Significant differences between the integral values in various
generalisations are observed. While $I_A$ and $I_B$ remain comparable at the
measured $Q^2$ values due to the smallness of $\gamma^2$, the 
$\frac{1}{1-x}$ weighting introduced by the Hand notation in $I_C$ leads to
sizeable differences.  The results will be discussed referring to the
generalisation $I_B$.

\begin{figure}
\centerline{\epsfig{figure=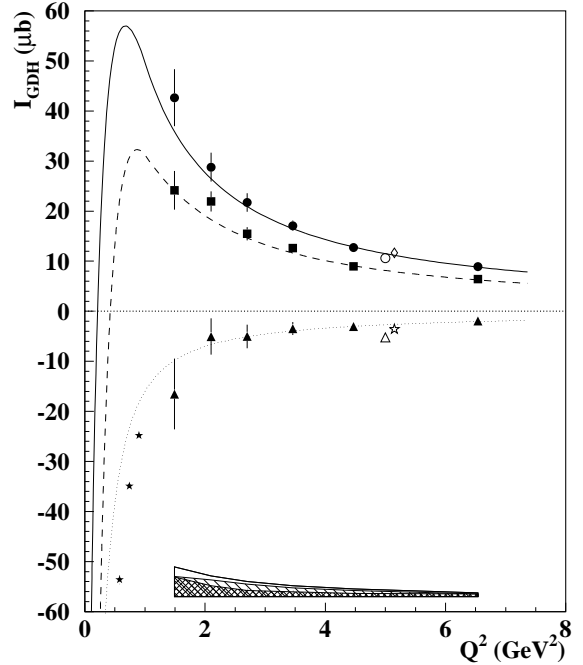,width=8.0cm}}
\caption{The $Q^2$--dependence of the generalised GDH integrals, calculated over the 
  full $W^2$--region, for the deuteron nucleus (squares), proton (circles) and neutron (triangles).
  The latter was obtained from the deuteron and proton data. The curves shown are the
  predictions for the various targets according to Ref.~\cite{teryaev}. 
  The error bars represent the statistical uncertainties. The bands represent the 
  systematic uncertainties (open: neutron, lined: deuteron, cross--hatched:
  proton). The open symbols at $Q^2=5\mbox{ GeV}^2$ represent the measurements from
  Ref.~\cite{SLAC}~(shifted to the left) and Ref.~\cite{E155}~(shifted to the
  right) on proton and neutron. The stars represent the three highest $Q^2$ bins of
  the neutron measurement from Ref.~\cite{hallA} including an extrapolation for the unmeasured DIS region.}
\label{fig:gdhtot}       
\end{figure}       

As mentioned above, the final results for the proton and for the neutron, as presented in Fig.~\ref{fig:proton}
and Fig.~\ref{fig:neutdis},  show that the contribution of the nucleon resonance region to the
full generalised GDH integral is small for $Q^2>3$ ~GeV$^2$ and the contribution from the 
DIS region remains sizeable down to the lowest measured $Q^2$
values. Numerical values following the generalisation $I_B$ are given in Table~\ref{tab:results1}.

The results for the full generalised GDH integrals on the deute\-ron, proton and
neutron are shown together in Fig.~\ref{fig:gdhtot}.
As noted above, they agree well within the total uncertainties
with a prediction of Ref.~\cite{teryaev} based on the leading twist $Q^2$--evolution of the two
polarised structure functions $g_1$ and $g_2$ without consideration of any
explicit nucleon--resonance contribution. Although in the neutron case the poorer knowledge of input data
for this model leads to a larger uncertainty, a similar description of the data
compared to the proton case is achieved. 
No turn--over is observed  in the measured range that would be required for the generalised GDH integral on the 
proton or the deuteron to meet the GDH sum rule predictions at $Q^2=0$. Preliminary data from Ref.~\cite{GDHJlab,CLAS} and a recent 
theoretical evaluation indicate that  this sign change happens at a
value of $Q^2$ much lower than the range considered in this analysis~\cite{ji}.

At large $Q^2$ the generalised GDH integral  is connected to the first moment of the
spin structure function $g_1$ (Eq. (\ref{eq:dis})).
 For $Q^2 > 3$ ~GeV$^2$ the generalised GDH integral is completely
dominated by the DIS region.
 The data presented in this paper agree  with
the most recent values for the first moments of the spin structure functions
measured on the proton by E--155 (E--143)
$\Gamma_1^p = 0.118 \pm 0.004 \pm 0.007 \quad (\Gamma_1^p=0.129 \pm 0.003 \pm 0.010) $
and on the neutron measured by E--155 (E--143)
$\Gamma_1^n=-0.058 \pm 0.005 \pm 0.008 \quad (\Gamma_1^n=-0.034 \pm 0.007 \pm 0.016) $
evaluated at $Q^2=5\mbox{~GeV}^2$~\cite{SLAC,E155}. These values correspond to
$I_{GDH}^p(Q^2=5\mbox{~GeV}^2)=10.59 \pm 0.36 \pm 0.73 \quad (11.85\pm 0.28
\pm 0.92)\; \mu b$ and
$I_{GDH}^n(Q^2=5\mbox{~GeV}^2)=-5.21 \pm 0.45 \pm 0.72 \quad (-3.12 \pm 0.64
\pm 1.47) \;\mu b$,
for the proton and the neutron respectively, and are shown together with the present data in
Fig.~\ref{fig:gdhtot}.

The $Q^2$--behaviour of $I_{GDH}$ for proton and neutron can be more clearly
studied when dividing out from $I_{GDH}$ the $1/Q^2$--dependence expected from leading twist. According to Eq.~(\ref{eq:dis}), $I_{GDH}$ is then expected to show a  logarithmic $Q^2$--dependence, similar to $\Gamma_1$. 
The result is shown in Fig.~\ref{fig:neutq2}. Any contributions from resonance form
factors or higher--twist contributions should become visible as a deviation from a
flat line. Considering the statistical and systematic uncertainties of the present
measurement, no deviation from a leading--twist behaviour can be seen. In other words, 
the experimental data obtained for the proton and the neutron are consistent with
the naive expectation that the $1/Q^2$ expansion is a good approximation down
to $Q^2 \simeq 2 {\mbox{~GeV}}^2$. As discussed in the introduction the elastic
part ($x=1$) has to be included for a complete comparison to a twist expansion of $\Gamma_1$. However, this is not a relevant contribution in
the kinematic range considered.

\begin{figure}
\centerline{\epsfig{figure=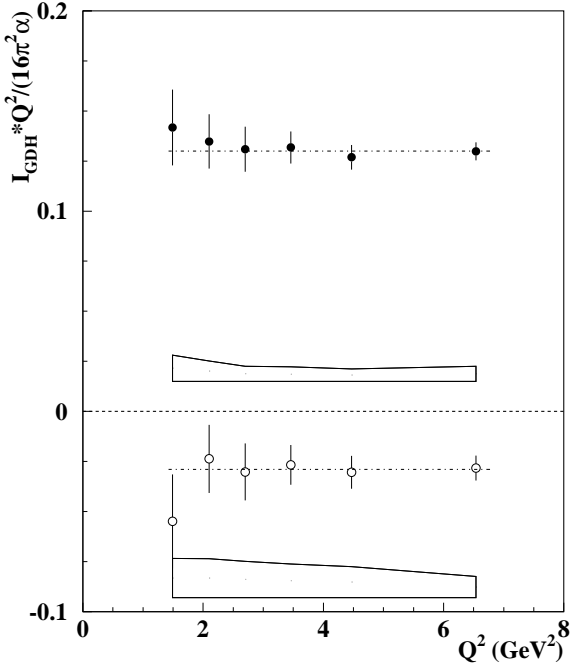,width=8.0cm}}
\caption{The $Q^2$--dependence of the generalised GDH integrals for the proton 
  (filled circles) and  neutron(open circles) 
  after the leading--twist dependence, $Q^2/(16\pi^2\alpha)$, has been divided out. 
  The error bars represent the
  statistical errors. The systematic uncertainties are represented by the
  respective error bands. The dash--dotted lines are straight line fits to the data.} 
\label{fig:neutq2}       
\end{figure}       

\begin{figure}
\centerline{\epsfig{figure=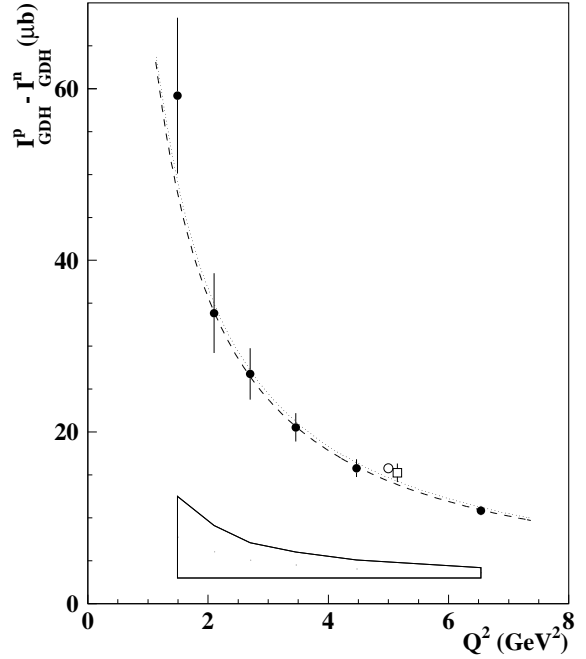,width=8.0cm}}
\caption{The $Q^2$--dependence of the generalised GDH integral for the proton--neutron
  difference. The dotted curve represent the prediction from Ref.~\cite{tertwo}. The dashed curve represents a simple $1/Q^2$ fit to the data leading to
  $I_{GDH}^{p-n}(Q^2=5\mbox{~GeV}^2)$ of 14.3 $\pm 0.9 \pm 1.3 \; \mu b$. For large $Q^2$ this difference is expected to obey
  the Bjorken sum rule.  The open symbols represent the measurements of
  the Bjorken sum rule from Refs.~\cite{SLAC} (square) and \cite{E155}
  (circle). They
  are slightly shifted in $Q^2$ for clearer representation. At $Q^2=0$ the GDH sum rule
  gives a value of 29 $\mu$b. The error bars represent the
  combined statistical uncertainties; the band indicates the
  systematic uncertainty.}
\label{fig:bjorken}       
\end{figure}

\begin{table*}
\caption{The generalised GDH integral $I_B$ for the deuteron, proton and neutron given
  in $\mu b$ per nucleus for the full $W^2$--region, the nucleon resonance region and the DIS region
   and various values of
  $Q^2$ in GeV$^2$. The statistical and systematic uncertainties are given. The last three rows show the integral values for the extrapolation to higher
   $W^2$, for the deuteron, proton and neutron. Note that the systematic error
   on the total integral is reduced compared to the integral in the nucleon-resonance
   and DIS regions, since smearing between these regions is not to be taken
   into account.}
\label{tab:results1}
\begin{center}
\begin{tabular}{|l|rcrcrcrcrcrc|}
\hline\noalign{\smallskip}
$Q^2$ & &\hspace{-1cm}$1.5$& &\hspace{-1cm}$2.1 $ & &\hspace{-1cm}$2.7 $ & &\hspace{-1cm}$3.5 $ & &\hspace{-1cm}$4.5 $ & &\hspace{-1cm}$6.5$ \\
\hline\noalign{\smallskip}\hline\noalign{\smallskip}
$I_{\rm tot}^d$ & $24.2$&\hspace{-0.45cm}$ \pm 3.9 \pm 3.9 $& $ 21.9$&\hspace{-0.45cm}$\pm 2.0 \pm 3.2 $& $
15.5$&\hspace{-0.45cm}$\pm 1.3 \pm 2.4 $& $12.6$&\hspace{-0.45cm}$\pm 0.7
\pm 1.7 $& $9.0$&\hspace{-0.45cm}$\pm 0.5 \pm 1.2 $& $6.4$&\hspace{-0.45cm}$\pm 0.3 \pm 0.5 $\\
$I_{\rm res}^d$ & $11.4$&\hspace{-0.45cm}$\pm 3.6 \pm 4.0 $& $ 9.3$&\hspace{-0.45cm}$\pm 1.8 \pm 3.3 $&
$4.2$&\hspace{-0.45cm}$\pm 1.1 \pm 2.2 $& $3.4$&\hspace{-0.45cm}$\pm 0.5 \pm 1.6 $& $0.9$&\hspace{-0.45cm}$\pm 0.3 \pm 0.9 $&
$0.1$&\hspace{-0.45cm}$\pm 0.1 \pm 0.1 $\\
$I_{\rm DIS}^d$ & $12.9$&\hspace{-0.45cm}$\pm 1.3 \pm 1.3 $& $ 12.3$&\hspace{-0.45cm}$\pm 1.0 \pm 1.3 $& $
10.6$&\hspace{-0.45cm}$\pm 0.8 \pm 1.3 $& $8.3$&\hspace{-0.45cm}$\pm 0.5 \pm 0.9$& $6.9$&\hspace{-0.45cm}$\pm 0.4 \pm 0.7 $&
$4.8$&\hspace{-0.45cm}$\pm 0.3 \pm 0.5 $\\
\hline\noalign{\smallskip}\hline\noalign{\smallskip}
$I_{\rm tot}^p$ & $42.7$&\hspace{-0.45cm}$\pm 5.7 \pm 3.8 $& $28.8$&\hspace{-0.45cm}$\pm 2.9 \pm 2.1 $&
$21.7$&\hspace{-0.45cm}$\pm 1.9 \pm 1.2 $& $17.1$&\hspace{-0.45cm}$\pm 1.0\pm 0.9 $& $12.7$&\hspace{-0.45cm}$\pm 0.6\pm 0.6 $&
$8.9$&\hspace{-0.45cm}$\pm 0.3 \pm 0.5 $\\
$I_{\rm res}^p$ & $22.2$&\hspace{-0.45cm}$\pm 5.4 \pm 4.2 $& $10.7$&\hspace{-0.45cm}$\pm 2.6  \pm 2.1 $&
$5.0$&\hspace{-0.45cm}$\pm 1.5 \pm 0.9 $& $2.5$&\hspace{-0.45cm}$\pm 0.7 \pm 0.4 $& $1.3$&\hspace{-0.45cm}$\pm 0.4 \pm 0.2 $&
$0.1$&\hspace{-0.45cm}$\pm 0.1 \pm 0.1 $\\
$I_{\rm DIS}^p$ & $17.2$&\hspace{-0.45cm}$\pm 1.9 \pm 1.1 $& $14.6$&\hspace{-0.45cm}$\pm 1.3 \pm 1.0 $&
$13.2$&\hspace{-0.45cm}$\pm 1.0 \pm 1.0 $& $11.1$&\hspace{-0.45cm}$\pm 0.7 \pm 0.8$& $7.9$&\hspace{-0.45cm}$\pm 0.5 \pm 0.6 $&
$5.5$&\hspace{-0.45cm}$\pm 0.3 \pm 0.5 $\\
\hline\noalign{\smallskip}\hline\noalign{\smallskip}
$I_{\rm tot}^n$ & $-16.5$&\hspace{-0.45cm}$\pm 7.1 \pm 5.7 $& $-5.1$&\hspace{-0.45cm}$\pm 3.6 \pm 4.0 $&
$-5.0$&\hspace{-0.45cm}$\pm 2.4 \pm 2.9 $& $-3.5$&\hspace{-0.45cm}$\pm 1.3 \pm 2.0 $& $-3.0$&\hspace{-0.45cm}$\pm 0.8 \pm 1.4
$& $-1.9$&\hspace{-0.45cm}$\pm 0.4 \pm 0.7 $\\
$I_{\rm res}^n$ & $-9.8$&\hspace{-0.45cm}$\pm 6.7 \pm 6.1 $& $-0.6$&\hspace{-0.45cm}$\pm 3.2 \pm 4.0 $&
$-0.6$&\hspace{-0.45cm}$\pm 1.9 \pm 2.6 $& $1.2$&\hspace{-0.45cm}$\pm 0.9 \pm 1.7 $& $-0.3$&\hspace{-0.45cm}$\pm 0.5 \pm 1.0 $&
$0.1$&\hspace{-0.45cm}$\pm 0.1 \pm 0.2 $\\
$I_{\rm DIS}^n$ & $-3.3$&\hspace{-0.45cm}$\pm 2.3 \pm 1.8$& $-1.3$&\hspace{-0.45cm}$\pm 1.7 \pm 1.7 $&
$-1.6$&\hspace{-0.45cm}$\pm 1.4 \pm 1.7 $& $ -2.1$&\hspace{-0.45cm}$\pm 0.9 \pm 1.3 $& $-0.4$&\hspace{-0.45cm}$\pm 0.6 \pm 1.0
$& $-0.2$&\hspace{-0.45cm}$\pm 0.4 \pm 0.7 $\\

\hline\noalign{\smallskip}
$I_{\rm unm}^d$ & $-0.1$&& $0.3$&&
$0.7$&& $0.9$&& $1.2$&&
$1.5$&\\
$I_{\rm unm}^p$ & $3.3$&& $3.5$&&
$3.5$&& $3.5$&& $3.5$&&
$3.3$& \\
$I_{\rm unm}^n$ & $-3.4$&& $-3.2$&&
$-2.8$&& $-2.6$&& $-2.3$&&
$-1.8$& \\
\hline\noalign{\smallskip}
\end{tabular}
\end{center}
\end{table*}

\begin{table*}
\caption{The generalised GDH integral for the deuteron, proton and neutron,
  calculated for the full $W^2$--range, given
  in $\mu b$ per nucleus for the  generalisations given in Eqs.~\ref{ia},
  \ref{ib}, \ref{ic} following Ref.~\cite{drechsel} and various values of
  $Q^2$ in GeV$^2$. The statistical and systematic uncertainties are given.}

\label{tab:results}
\begin{center}
\begin{tabular}{|l|rcrcrcrcrcrc|}
\noalign{\smallskip}\hline
$Q^2$ & &\hspace{-1cm}$1.5$& &\hspace{-1cm}$2.1 $ & &\hspace{-1cm}$2.7 $ &
&\hspace{-1cm}$3.5 $ & &\hspace{-1cm}$4.5 $ & &\hspace{-1cm}$6.5$ \\
\noalign{\smallskip}\hline\noalign{\smallskip}\hline
$I_A^d $ & $27.4 $ & \hspace{-0.45cm} $\pm 4.7 \pm 4.5 $ & $ 24.5 $ &
\hspace{-0.45cm} $\pm 2.4 \pm 3.5 $ & $16.7 $ & \hspace{-0.45cm} $\pm 1.6 \pm
2.6 $ & $13.5 $ & \hspace{-0.45cm} $\pm 0.8 \pm 1.8 $ &
 $ 9.4 $ & \hspace{-0.45cm} $\pm 0.5  \pm 1.2 $& $ 6.6$ & \hspace{-0.45cm}
 $\pm 0.3   \pm 0.5 $ \\
$I_B^d $ & $ 24.2 $ & \hspace{-0.45cm} $\pm 3.9  \pm 3.9  $ & $  21.9 $ &
\hspace{-0.45cm} $\pm 2.0  \pm 3.2  $ & $ 15.5 $ & \hspace{-0.45cm} $\pm 1.3
\pm 2.4
 $ & $ 12.6 $ & \hspace{-0.45cm} $\pm 0.7 \pm 1.7  $ &
 $ 9.0 $ & \hspace{-0.45cm} $\pm 0.5 \pm  1.2 $& $ 6.4 $ & \hspace{-0.45cm}
 $\pm 0.3   \pm 0.5  $ \\
$I_C^d $ & $ 44.9 $ & \hspace{-0.45cm} $\pm 9.3  \pm 7.3  $ & $ 43.3 $ &
\hspace{-0.45cm} $\pm 5.2  \pm 6.3  $ & $ 28.3 $ & \hspace{-0.45cm} $\pm 3.6
\pm 4.4
 $ & $ 24.5 $ & \hspace{-0.45cm} $\pm 1.9 \pm 3.3  $ &
 $ 14.9 $ & \hspace{-0.45cm} $\pm 1.3  \pm 2.0  $& $ 9.4 $ & \hspace{-0.45cm}
 $\pm 0.4    \pm 0.8 $ \\
\noalign{\smallskip}\hline\noalign{\smallskip}\hline
$I_A^p $ & $ 48.7 $ & \hspace{-0.45cm} $\pm 6.9  \pm 4.3  $ & $ 31.7  $ &
\hspace{-0.45cm} $\pm 3.4  \pm 2.3  $ & $ 23.2 $ & \hspace{-0.45cm} $\pm 2.2
\pm 1.3
 $ & $17.9 $ & \hspace{-0.45cm} $\pm 1.1  \pm 1.0  $ &
 $ 13.3 $ & \hspace{-0.45cm} $\pm 0.7  \pm 0.6  $& $ 9.0 $ & \hspace{-0.45cm}
 $\pm 0.3   \pm 0.5 $ \\
$I_B^p $ & $42.7 $ & \hspace{-0.45cm} $\pm 5.7  \pm 3.8 $ & $ 28.8 $ &
\hspace{-0.45cm} $\pm 2.9  \pm 2.1 $ & $21.7 $ & \hspace{-0.45cm} $\pm 1.9
\pm 1.2
 $ & $ 17.1 $ & \hspace{-0.45cm} $\pm 1.0  \pm 0.9 $ &
 $ 12.7 $ & \hspace{-0.45cm} $\pm 0.6  \pm 0.6  $& $ 8.9 $ & \hspace{-0.45cm}
 $\pm 0.3    \pm 0.5  $ \\
$I_C^p $ & $ 81.7 $ & \hspace{-0.45cm} $\pm 14.0  \pm 7.4  $ & $ 53.4 $ &
\hspace{-0.45cm} $\pm 7.7  \pm 3.9  $ & $ 37.4 $ & \hspace{-0.45cm} $\pm 5.3
\pm 2.1
 $ & $ 27.8 $ & \hspace{-0.45cm} $\pm 2.8  \pm 1.6 $ &
 $ 20.4 $ & \hspace{-0.45cm} $\pm 1.9 \pm 0.9  $& $ 11.8 $ & \hspace{-0.45cm}
 $\pm 0.4   \pm 0.7  $ \\
\noalign{\smallskip}\hline\noalign{\smallskip}\hline
$I_A^n $ & $-19.0 $ & \hspace{-0.45cm} $\pm 8.6  \pm 6.5  $ & $ -5.2 $ &
\hspace{-0.45cm} $\pm 4.3  \pm 4.5  $ & $ -5.1 $ & \hspace{-0.45cm} $\pm 2.8 \pm
3.1 $ & $-3.3 $ & \hspace{-0.45cm} $\pm 1.4  \pm 2.2  $ &
 $ -3.1 $ & \hspace{-0.45cm} $\pm 0.9  \pm 1.5  $& $ -1.9 $ & \hspace{-0.45cm}
 $\pm 0.4    \pm 0.7 $ \\
$I_B^n $ & $ -16.5 $ & \hspace{-0.45cm} $\pm 7.1  \pm 5.7  $ & $ -5.1  $ &
\hspace{-0.45cm} $\pm 3.6  \pm 4.0 $ & $-5.0 $ & \hspace{-0.45cm} $\pm 2.3
\pm 2.9
 $ & $ -3.5 $ & \hspace{-0.45cm} $\pm 1.3  \pm 2.0  $ &
 $ -3.0 $ & \hspace{-0.45cm} $\pm 0.8  \pm 1.4  $& $ -2.0 $ & \hspace{-0.45cm}
 $\pm 0.4   \pm 0.7  $ \\
$I_C^n $ & $ -33.1 $ & \hspace{-0.45cm} $\pm 17.2  \pm 10.8 $ & $ -6.6  $ &
\hspace{-0.45cm} $\pm 9.5  \pm 7.8  $ & $ -6.6 $ & \hspace{-0.45cm} $\pm 6.6
\pm 5.1
 $ & $-1.4 $ & \hspace{-0.45cm} $\pm 3.5  \pm 3.8  $ &
 $ -4.3 $ & \hspace{-0.45cm} $\pm 2.3  \pm 2.3  $& $ -1.6 $ & \hspace{-0.45cm}
 $\pm 0.6    \pm 1.0  $ \\
\noalign{\smallskip}\hline
\end{tabular}
\end{center}
\end{table*}

\section{Combined results for the Proton and the Neutron}\label{sec:bsr}
The data obtained for the proton and neutron  over a large range in
$Q^2$ and $W^2$ offer a unique possibility to evaluate the proton--neutron
difference of the generalised GDH integral. This difference is shown in
Fig.~\ref{fig:bjorken}. It is expected to be less sensitive to higher twist effects or
contributions from nucleon resonances.
Within the measured $Q^2$--range no turn--over at low $Q^2$  
 required to meet the GDH sum rule prediction of 29 $\mu b$ for $Q^2=0$ is observed. The data fall off as $1/Q^2$ indicating that leading twist
dominates as expected. A fit to the data using $c/Q^2$ where $c$ is a constant is shown
together with the data in Fig.~\ref{fig:bjorken}. For the  constant $c/(16\pi^2\alpha)= 0.159 \pm 0.009$ is
found, leading to a value $I_{GDH}^{p}-I_{GDH}^{n}= 14.3\pm 0.9 \pm
1.3 \; \mu b$  at $Q^2=5 \mbox{~GeV}^2$. This result is  in agreement  with an experimental
determination  of the Bjorken sum rule by E--155 (E--143), $ \Gamma_1^p-\Gamma_1^n=0.176 \pm
0.003 \pm 0.007 \quad (0.164 \pm 0.008 \pm 0.020)$ ~\cite{SLAC,E155} within
the respective experimental uncertainties. At
$Q^2=5\mbox{~GeV}^2$ these values correspond to $ I_{GDH}^{p}-I_{GDH}^{n}=15.76 \pm 0.27 \pm 0.63\; \mu
b \quad (14.95 \pm 0.72\pm 1.79\; \mu b)$.
Within errors, the value for the proton--neutron difference measured in this analysis is also 
in agreement with the Bjorken--sum--rule prediction of 
$0.182 \pm 0.005$ corresponding to $I_{GDH}^{p}-I_{GDH}^{n}=16.33\pm 0.45\; \mu b$. The Bjorken sum rule
was  evaluated using Eq.~(\ref{eq:bjorken})  at
$Q^2=5 \mbox{~GeV}^2$ including corrections up to third--order
in $\alpha_s$ and with $\alpha_s(M_Z)=0.118$.

\section{Summary}\label{sec:sum}
The generalised GDH integral has been determined for a deuteron target in the
 kinematic region  $1.2 < Q^2 < 12.0$ ~GeV$^2$ and $1 < W^2<45 $ ~GeV$^2$
 in this paper. Using an updated value of the target polarisation, a corresponding re--analysis 
of the HERMES proton data as compared to Ref.~\cite{gdhres} was performed.
In both cases the $W^2$--range was separated at $W^2 = 4.2$ ~GeV$^2$ 
into a region where the nucleon resonances dominate and into the DIS region. 
Combining both data sets, the generalised GDH integral for the neutron was calculated in the 
same kinematic regions. These neutron results obtained from the deuteron agree with those
obtained earlier on a $^3$He target in the same kinematic region.

Altogether, a complete set of measurements of the generalised GDH integrals for the
deuteron, proton and neutron is available. In all three cases
the nucleon-resonance contribution to the generalised integral decreases rapidly with increasing
$Q^2$ and the contribution from the DIS region is still sizeable
even at the lowest measured $Q^2$, emphasising the importance of measuring the
GDH integral over a large $W^2$--range.
At larger $Q^2$ the measured values agree well with measurements of the first moments of the
spin structure function $g_1$. 

For the generalised GDH integrals of the proton, the neutron
and the proton--neutron difference, the $Q^2$--de\-pend\-ence is
in agreement with a leading--twist behaviour; within the experimental uncertainties
it exhibits no significant contribution from either higher twist or resonance form factors. 
The proton--neutron difference is in agreement with the Bjorken--sum--rule
prediction evaluated at $Q^2=5\mbox{~GeV}^2$ within the experimental uncertainties.

\begin{acknowledgement}
We gratefully acknowledge the DESY management for its support, the staffs at
DESY and the collaborating institutions for their significant effort.
This work was supported by the FWO--Flanders, Belgium; the Natural Sciences
and Engineering Research Council of Canada; the INTAS and RTN network ESOP (contract number 1999-00117)
contributions from the European Union; the German Bundesministerium f\"ur
Bildung und Forschung; the Deutsche For\-schungs\-gemeinschaft (DFG); the
Deutscher Akademischer Aus\-tausch\-dienst (DAAD); the Italian Istituto Nazionale
di Fisica Nucleare (INFN); Monbusho International Scientific Research Program,
JSPS, and Toray Science Foundation of Japan; the Dutch Foundation for
Fundamenteel Onderzoek der Materie (FOM); the U.~K. Particle Physics and
Astronomy Research Council; and the U.~S. Department of Energy and National
Science Foundation.
\end{acknowledgement}



\begin{thebibliography}{}
\bibitem{ger} S.B.~Gerasimov, Sov. J. Nucl. Phys. {\bf 2} (1966) 430.

\bibitem{dh} S.D.~Drell and A.C.~Hearn, Phys. Rev. Lett. {\bf 16} (1966) 908.

\bibitem{electroweak} G.~Altarelli, N.~Cabbibo, M.~Maiani, Phys. Lett. {\bf B
    40} (1972) 415; S.~Brodsky, I.~Schmidt, Phys. Lett. {\bf B 352} (1995) 344.

\bibitem{MAMIres} GDH collaboration,  J.~Ahrens {\em et al.}, Phys. Rev. Lett. {\bf 87} (2001)
  022003.

\bibitem{ELSA} GDH collaboration, G.~Zeitler, $\pi$N--Newsletter {\bf 16} (2002) 311.

\bibitem{otherexp} V. Ghazikhanian {\em et al.}, SLAC--Proposal E--159 (2000);
  D.Sober {\em et al.}, CEBAF PR--91--15 (1991).

\bibitem{RP} R.~Pantf\"order, PhD Thesis, Universit\"at Bonn (1998), BONN-IR-98-06, hep-ph/9805434 and references therein.

\bibitem{drechsel} D.~Drechsel, S.S.~Kamalov and L.~Tiator, Phys. Rev. {\bf D
    63} (2001), 114010.

\bibitem{JiMel} X.~Ji and W.~Melnitchouk, Phys. Rev. {\bf D 56} (1997) R1.

\bibitem{RP567} I.~Karliner, Phys. Rev. {\bf D 7} (1973) 2717;
R.L.~Workman and R.A.~Arndt, Phys. Rev. {\bf D 45} (1992) 1789;
A.M.~Sandorfi, C.S.~Whisnant and M.~Khandaker, Phys. Rev. {\bf D 50} (1994)
 R6681;
D.~Drechsel and G.~Krein, Phys. Rev. {\bf D 58} (1998) 116009.

\bibitem{bjorken} J.D.~Bjorken, Phys. Rev. {\bf 148} (1966) 1467;
  Phys. Rev. {\bf D 1} (1970) 1376.

\bibitem{PDG} Particle Data Group, Eur. Phys. J. {\bf C 15} (2000) 1.

\bibitem{lar2} S.~A.~Larin, J.~A.~M.~Vermaeseren, Phys. Lett. {\bf B 259}
  (1991) and references therein.


\bibitem{SLAC} E--143 Collaboration, K.~Abe {\em et al.}, Phys. Rev. {\bf D 58}
 (1998) 112003. 

\bibitem{E155} E--155 Collaboration, P.~L.~Anthony {\em et al.}, Phys. Lett. {\bf
    B 493} (2000) 19; E--155 Collaboration, P.~L.~Anthony {\em et al.},
    Phys. Lett. {\bf  B 463} (1999) 339.

\bibitem{SMC} SMC Collaboration, B.~Adeva {\em et al.}, Phys. Rev. {\bf D 58}
  (1998)  112002.

\bibitem{gil} F.J.~Gilman, Phys. Rev. {\bf 167} (1968) 1365.

\bibitem{Hand} L.~N.~Hand, Phys. Rev. {\bf 129} (1963) 1834.

\bibitem{GDHJlab} V.~Burkert {\em et al.}, CEBAF PR-91-23, 1991; S.~Kuhn {\em et
    al.}, CEBAF PR-93-09, 1993;  J.P.~Chen {\em et al.}, TJANF PR-97-110, 1997.

\bibitem{CLAS} CLAS Collaboration, R.~Fatemi {\em et al.}, Proc. of SPIN 2000: 14th International Spin Physics Symposium, edited by K.~Hatanaka {\em et al.}, Osaka, Japan, AIP Conf. Proc. {\bf 570} (2001) 402.

\bibitem{hallA} M.~Amarian {\em et al.}, {\tt arXiv:nucl-ex/0205020}, Phys. Rev.Lett. in press; priv. comm.

\bibitem{gdhdis} HERMES Collaboration, K.~Ackerstaff {\em et al.}, Phys. Lett.
{\bf B 444} (1998) 531.

\bibitem{gdhres} HERMES Collaboration, A. Airapetian {\em et al.},
  Phys. Lett. {\bf B 494} (2000) 1.

\bibitem{bar} D.P.~Barber {\em et al.}, Phys. Lett. {\bf B 343} (1997) 436.

\bibitem{beck} M. Beckmann {\em et al.}, Nucl. Instr. Meth. {\bf A 479} (2002) 334.

\bibitem{ste} J.~Stewart, Proc. of the Workshop on Polarised gas targets and
polarised beams, edited by R.J.~Holt and M.A.~Miller, Urbana-Champaign, USA,
AIP Conf. Proc. {\bf 421} (1997) 69.   

\bibitem{sto} F.~Stock {\em et al.}, Nucl. Instr. and Meth. {\bf A 343} (1994) 334.

\bibitem{BRP} C.~Baumgarten {\em et al.}, Nucl. Instr. and Meth.{\bf A 482} (2002) 606.


\bibitem{TGA} M.~C.~Simani, PhD Thesis, Vrije Universtiteit Amsterdam, October 2002. 

\bibitem{targetpaper} HERMES Collaboration, P. Lenisa, talk at the 15th Int. Spin Physics Symposium, Upton/USA, Sept. 2002, to appear in the proceedings

\bibitem{lumi} Th.~Benisch {\em et al.}, Nucl. Instr. and Meth. {\bf A 471} (2001) 314.

\bibitem{hspect} HERMES Collaboration, K.~Ackerstaff {\em et al.}, Nucl. Instr.
and Meth. {\bf A 417} (1998) 230.

\bibitem{g1p} HERMES Collaboration, A.~Airapetian {\em et al.},
Phys. Lett. {\bf B 442} (1998) 484.

\bibitem{RICH} N. Akopov et al., Nucl. Instr. and Meth. {\bf A 479} (2002) 511.

\bibitem{polrad} I.V.~Akushevich and N.M.~Shumeiko, J. Phys. {\bf G 20} (1994)
 513; I.~Akushevich {\em et al.}, Comput. Phys. Commun. {\bf 104} (1997) 201.

\bibitem{nmcd8}  NMC Collaboration, M.~Arneodo {\em et al.}, Nucl. Phys. {\bf B
    371} (1992) 3.

\bibitem{nmc} NMC Collaboration, M.~Arneodo {\em et al.}, Phys. Lett. {\bf B 36
4} (1995) 107.

\bibitem{stein} S.~Stein {\em et al.}, Phys. Rev. {\bf D 12} (1975) 1884 and
  references therein.

\bibitem{bilen} S.~I.~Bilen'kaya {\em et al.}, Zh. Eksp. Teor. Fiz. Pis'ma {\bf
    19} (1974) 613.

\bibitem{bodekneut} A.~Bodek, Phys. Rev. {\bf D 8} (1973) 2331.

\bibitem{nagaitsev} A.~P.~Nagaitsev {\em et al.}, JINR Rapid Communications,
  July 1995, N3(71)-95, 59.

\bibitem{ede} J.~Edelmann, G.~Piller, N.~Kaiser and W.~Weise, Nucl. Phys. {\bf
    A 665} (2000) 125.

\bibitem{whi} L.W.~Whitlow {\em et al.},  Phys. Lett. {\bf B 250} (1990) 193.

\bibitem{nikolo} N.~Bianchi and E.~Thomas, Phys. Lett. {\bf B 450} (1999) 439.

\bibitem{azna} I.~G.~Aznauryan, Phys. of At. Nucl. {\bf 58} (1995) 1014
and private communication.

\bibitem{teryaev}
J.~Soffer and O.V. ~Teryaev, Phys. Rev. {\bf D 51} (1995) 25;
J.~Soffer and O.V.~Teryaev, Phys. Rev. Lett. {\bf 70} (1993) 3373.

\bibitem{tertwo} J.~Soffer and O.V.~Teryaev, Phys.Rev. {\bf D 56} (1997) 7458;
  J.~Soffer and O.V.~Teryaev, hep--ph/0207252.

\bibitem{ciofi} C. Ciofi degli Atti {\em et al.}, Phys. Lett. {\bf B 376}
  (1996) 309.


\bibitem{lacombe} M.~Lacombe {\em et al.}, Phys. Rev. {\bf C 21} (1980) 861;
  M.~J.~Zuilhof, J.~A.~Tjon, Phys. Rev. {\bf C 22} (1980) 2369; A.~Yu.~Umnikov
  {\em et al.}, proceedings of SPIN 1994: 11th Int. Symp. on High Energy Spin Physics, edited by K.J.~Heller, S.L.~Smith , AIP Conf. Proc. {\bf 343} 79. 

\bibitem{arenhoevel} H.~Arenh\"ovel, G.~Kre\ss, R.~Schmidt, P.~Wilhelm,
  Phys. Lett. {\bf B 407} (1997) 1.

\bibitem{terthree} O.V.~Teryaev, talk at XVI
International Baldin Seminar "Relativistic Nuclear Physics and Quantum
Chromodynamics", Dubna/Russia, June 2002, to appear in the proceedings.

\bibitem{deltaq} HERMES Collaboration, K.~Ackerstaff {\em et al.},
  Phys. Lett.{\bf B 464} (1999) 123; HERMES Collaboration, M.~Beckmann,
  Proc. of the Workshop on "Testing QCD through Spin Observables on Nuclear
  Targets", Charlottesville, VA, 2002


\bibitem{ji} X.~Ji, C.W.~Kao and J.~Osborne, Phys.Rev. {\bf D 61} (2000) 074003
; X.~Ji and J.~Osborne, J. Phys. {\bf G 27} (2001) 127.

\end{thebibliography}
\end{document}